\documentclass[a4paper,10pt,twoside]{cpc-hepnp}
\usepackage{CJK,upgreek,fancyhdr}
\usepackage{multicol}
\usepackage{graphicx}
\usepackage{booktabs}
\usepackage{amssymb,bm,mathrsfs,bbm,amscd}
\usepackage[tbtags]{amsmath}
\usepackage{lastpage}
\usepackage{epsfig}

\begin{document}
\begin{CJK*}{GB}{gbsn}

\fancyhead[c]{\small submitted to Chinese Physics C}


\title{Analytical description of the excited state phase transition  to octupole deformed shape in alternating parity bands
\thanks{Supported by the RFBR (Moscow) (Grants No. 16-02-00068A)
 and   the  Russian Government Subsidy Program of the Competitive Growth of Kazan Federal University. 
 S.G.Z. was partly supported by  the National Key R\&D Program of China (2018YFA0404400), the NSF of China (11525524, 11621131001, 11647601, 11747601, and 11711540016), the Key Research Program of Frontier Sciences of CAS,
and the IAEA CRP ``F41033'' and the HPC Cluster of SKLTP/ITP-CAS and the Supercomputing Center, CNIC of CAS.
}
}
\author{%
       E.V. Mardyban$^{1,2;1)}$\email{mardyban@theor.jinr.ru}%
\quad T.M. Shneidman $^{1,3;2)}$\email{shneyd@theor.jinr.ru}%
\quad E.A. Kolganova$^{1,2}$
\quad R.V. Jolos$^{1,2}$
\quad S.-G. Zhou$^{4,5,6,7}$
}
\maketitle

\address{%
$^1$ Bogoliubov Laboratory of Theoretical Physics, Joint Institute for Nuclear Research,  Dubna, 141980, Russia\\
$^2$ Dubna State University, Dubna 141980, Russia\\
$^3$ Kazan Federal University, Kazan 420008, Russia\\
$^4$ CAS Key Laboratory of Theoretical Physics,   Institute of Theoretical Physics, Chinese Academy of Sciences, Beijing 100190, China\\
$^5$ School of Physical Sciences, University of Chinese Academy of Sciences, Beijing 100049, China\\
$^6$ Center of Theoretical Nuclear Physics, National Laboratory  of Heavy Ion Accelerator, Lanzhou 730000, China\\
$^7$ Synergetic Innovation Center for Quantum Effects and Application,   Hunan Normal University, Changsha, 410081, China\\

}

\begin{abstract}
Angular momentum dependences of the parity splitting and electric dipole transitions in the alternating parity bands of heavy nuclei have been analyzed. It is shown that these dependences can be treated in a universal way with  a single parameter of critical angular momentum, which characterizes 
 phase transition from octupole vibrations  to the stable octupole deformation. Using the simple but useful model of axially-symmetric reflection-asymmetric mode, the analytical expression for the parity splitting and electric dipole transitional moment have been obtained. The results obtained are in a good agreement with the experimental data for various isotopes of Ra, Th, U, and Pu. 
\end{abstract}

\begin{keyword}
octupole deformation, phase transition, angular momentum
\end{keyword}

\begin{pacs}
21.10.Re, 21.10.Ky, 21.60.Ev
\end{pacs}

\footnotetext[0]{\hspace*{-3mm}\raisebox{0.3ex}{$\scriptstyle\copyright$}2013
Chinese Physical Society and the Institute of High Energy Physics
of the Chinese Academy of Sciences and the Institute
of Modern Physics of the Chinese Academy of Sciences and IOP Publishing Ltd}%

\begin{multicols}{2}

\section{Introduction}

Phase transitions in nuclei attract much attention in recent time as the nucleus provides plentiful amount of examples.  Mainly, the phase transitions between spherical, axially deformed, 
and $\gamma$-soft limits of nuclear structure have been analyzed \cite{Cejnar2010}. Recently, the evolution of reflection asymmetric deformation in actinides and rare-earth nuclei starts
to draw much attention as well. It shows an interesting case of the phase transition occurring with angular momentum. 
Indeed, as it was shown in \cite{Jolos2012},  the  evolution of the parity splitting in the alternating parity bands in actinides clearly demonstrates  the transition between octupole non-deformed phase to the stable octupole deformation.
This phenomenon can also be considered as an example of the excited state quantum phase transition \cite{Iachello}.

It is well known that many nuclei in the actinides and rare-earth mass regions are soft with respect to the deformations which violates the spatial reversal symmetry.  Experimentally it is revealed by the appearance of the low-lying negative parity states connected by the strong (collective) odd-multipolarity transitions with the members of the ground
 state band \cite{Butler1996}.  Since the first observation of low energy negative parity states \cite{Asaro1953,Asaro1955},  an extensive set of experimental data has been accumulated (for the review see \cite{Ahmad1993}). 
Note also the recent experimental investigations concerning the reflection-asymmetry in  $^{218,220}$Rn and $^{222,224}$Ra  \cite{Gaffney2013},  in  $^{240}$Pu \cite{Spieker2013}, in $^{143}$Ba \cite{Chen2016_PRC94-021301R} and in $^{144,146}$Ba \cite{Bucher2016}.

In nuclei with strong quadrupole deformation, yrast  negative parity states  constitute the rotational band  1$^-$, 3$^-$, 5$^-$ ... In the case of static reflection-asymmetric deformation, these states together
 with the members of the ground state band would form the unified band with the negative and positive parity states interleaved each other following the rotational order with equal moment of inertia. However, 
in most even-even nuclei, at low angular momenta, the negative parity states are shifted up with respect to the positions which they would have in a  unified  alternating parity band of molecular type. This shift denoted by the parity splitting \cite{Jolos1994} point out 
that at low angular momenta nuclei perform the vibrational dynamics  in reflection-asymmetric degree of freedom. Indeed, the results of calculations within the shell-corrected liquid drop models \cite{Nazarewicz1984, Moller2008} 
and mean-field models \cite{Egido1989, Rutz1995, Robledo2011,Lu2014,Zhou2016,Zhao2012} show that nuclei in these mass regions, although being soft with respect to the octupole deformation,  does not develop strong minimum at non zero values of reflection-asymmetric deformation.

Therefore, we see that although the reflection-asymmetric deformation is not as stable as conventional quadrupole deformation, it is very important for the description of the structure of the excitation spectra. 
  However, with the increase of angular momentum the parity splitting decreases and almost unperturbed alternating parity band is formed. That means that reflection-asymmetric deformation stabilizes. 
Therefore, with angular momentum, the  transition occurs from the reflection symmetric to reflection-asymmetric phase. In  \cite{Jolos2012}  it was shown that one can define the crtical angular momentum at which this transition occurs.

Another observable sensitive to the strength of the reflection-asymmetric deformation is the reduced transition probability for the electric dipole transitions between the states of the negative parity
  and the ground state band. Data on angular momentum dependence of the dipole moment are not as rich as for the energy spectra (for the review, see \cite{Butler1996}). 
Electric dipole transitional moment increases with angular momentum until some critical value after which it remains almost constant. Such a behavior of the dipole moment is consistent with the idea of the phase transition.  

It is interesting to note that in odd-mass nuclei the stabilization of the reflection-asymmetry occur earlier than in even-even neighbours.
 As shown in \cite{Leander2003}, an interplay between the single-particle motion and collective reflection-asymmetric degree of freedom leads to the reduction of the parity splitting.

Theoretical models developed to describe nuclear reflection-asymmetry dynamics  depends on what  degree of freedom is used. This degree of freedom is related either to the 
octupole deformation \cite{Nazmitdinov1987,Minkov2013} or to the clustering \cite{Shneidman2003,Buck1999,Iachello2}. In the frame of these models it is possible to obtain qualitative and quantitative 
description of evolution of reflection-asymmetric deformation with mass and charge number,  as well as the energies of the lowest negative parity excitations and their decay properties.   
In the frame of the    interacting boson model extended to include dipole and octupole bosons,  a nice reproduction of
experimental data is obtained \cite{Kusnezov1988}. In the cluster approach based on the semi-microscopical dinuclear system model, a satisfactory description of the parity splitting and $E\lambda$ transition probabilities  in many actinides has been obtained \cite{Shneidman2014}.
However, the evolution of reflection-asymmetry   is hard to analyze fully microscopically since it requires to perform the calculation up to a large values of angular momentum.  An attempt to carry out the  calculations in $^{144}$Ba
for the lowest positive and negative parity states has been performed in GCM framework with angular momentum, parity and particle number projected HFB wave functions \cite{Robledo2016}. 
The  same technique  has been applied to the description of the lowest states in $^{224}$Ra  with the relativistic mean field wave functions \cite{EFZhou2016}.

Despite of these difficulties the analysis of the experimental data shows that  the  behavior of the parity splitting and transitional dipole moment with angular momentum looks quite universal. 
In other words, basing on the general ideas about the reflection-asymmetric mode (regardless, octupole or mass-asymmetry)  one can propose simple analytical description of the angular momentum dependence of these
quantities containing a small number of the parameters having a clear physical meaning. These parameters can be further fitted to the experiment or calculated in the microscopic models.  
It is the aim of this paper to find such an analytical description of the angular momentum dependence of the main physical characteristics of the alternating parity bands.

\section{Description of the model}
\subsection{Hamiltonian and eigenfunctions}

Following  the discussion in \cite{Jolos2011}, we assume that the nucleus under consideration has a static quadrupole deformation $\langle\beta_{20} \rangle$ and is soft with respect to the axially-symmetric ($K=0$) reflection-asymmetric  vibrations.  These vibrations 
can be generated either by the octupole  or by the mass asymmetry degree of freedom.  In both cases we denote the corresponding dynamical variable by $\beta_{30}$, although in the case of the mass asymmetry degree of freedom, the contribution of the higher order odd multipolarity modes is effectively included. The intrinsic Hamiltonian describing 
collective motion in $\beta_{30}$ for a given angular momentum $I$ can be written as
\begin{eqnarray}
H_I=-\dfrac{\hbar^2}{2 B}\dfrac{d^2}{d \beta_{30}^2}+V_I(\langle\beta_{20} \rangle,\beta_{30}),
\label{hamiltonian}
\end{eqnarray}  
where $B$ is the effective mass. The potential energy $V_I$  is an even function of $\beta_{30}$.

The numerical diagonalization of the Hamiltonian (\ref{hamiltonian}) with different variants of the potential $V_I$ has shown \cite {Jolos2005,Shneidman2003} that with a good accuracy the lowest eigenstate of positive parity can be approximated as a superposition of two Gaussians 
of width  $\sqrt{\hbar/(B \omega(I))}$  centered at $\beta_{30}= \pm \beta_m(I)$
\end{multicols}
\begin{eqnarray}
\Psi^{(+)}_I(\beta_{30})&=&\left [ \dfrac{\omega(I)}{\pi \hbar}\right ]^{1/4} \left ( 2 \left \{ 1+\exp{\left [-\dfrac{B \omega(I)}{\hbar}\beta_m^2(I) \right ]}\right \}\right )^{-1/2} \nonumber \\
&\times& \left ( \exp{\left [-\dfrac{B \omega(I)}{2\hbar}(\beta_{30}-\beta_m(I))^2 \right ]} + \exp{\left [-\dfrac{B \omega(I)}{2\hbar}(\beta_{30}+\beta_m(I))^2 \right ]}\right ).
\label{ansatz}
\end{eqnarray}
\begin{multicols}{2}
Above, $\omega(I)$ is, in principle, a function of the angular momentum which is determined further.
The convenience of the ansatz (\ref{ansatz}) for the positive parity wave functions is related to the fact that both, the limit of octupole vibrations and the limit of stable octupole deformation are described equally well by this ansatz \cite{Jolos2011}.  One can introduce the parameter 
\begin{eqnarray}
\xi_I=\sqrt{ \dfrac{B \omega(I)}{\hbar}} \beta_m(I),
\label{xi}
\end{eqnarray}
which gives the ratio of a distance between the centers of the Gaussians to the sum of their  widths. If $\xi \ll 1$, the overlap of the components in (\ref{ansatz}) is large and the wave function $\Psi^{(+)}_I$ corresponds to the case of octupole vibrations. If $\xi \gg 1$, the situation is opposite and the components in (\ref{ansatz}) are well separated. The later corresponds to the static octupole deformation.

Having the wave function $\Psi^{(+)}_I$, one can obtain from the Schr\"odinger equation with the Hamiltonian (\ref{hamiltonian}) the potential for the axially-symmetric octupole mode as
\begin{eqnarray}
V_I(\langle \beta_{20}\rangle, \beta_{30})=\dfrac{\hbar^2}{2 B} \dfrac{\Psi^{(+)''}_I}{\Psi^{(+)}_I}+E_I^{(+)}(\langle \beta_{20}\rangle),
\label{potential_1}
\end{eqnarray}
where $E_I^{(+)}(\langle \beta_{20}\rangle)$ is the excitation energy of the lowest state with an angular momentum $I$ and the positive parity.  In this paper we are interested in the calculation of the parity splitting which is  determined as a difference between  the energies of the 
 negative parity $E_I^{(-)}$ and the positive parity  $E_I^{(+)}$ states having the same angular momentum $I$.
 Since, due to the $K=0$ selection rules,  there exists only one physical ecxited state  for a given $I$ (positive parity for even $I$ and negative parity for odd $I$),  an experimental parity splitting can be determined as a difference between the 
experimental excitation energy for one parity and the energy obtained by interpolation  between the energies of the neighboring states of the opposite parity (see \cite{Jolos1994} or Eq.~(\ref{interpolation}) of the present paper).
 Thus,  $E_I^{(+)}(\langle\beta_{20}\rangle)$   never enters the final results and we can set it equal to zero. Note, however, that as discussed in Sect.~(3.2), this can only be done for the well-deformed nuclei.

The ansatz (\ref{ansatz}) for the wave function $\Psi^{(+)}_I(\beta_{30})$ yields the following expression for the potential energy of the axially-symmetric reflection-asymmetric mode determined up to $I$-dependent constant

\begin{eqnarray}
V_I(\beta_{30})&=&\dfrac{\hbar \omega}{2} \left (-1+\dfrac{B \omega(I)}{\hbar}(\beta_{30}^2+\beta_m^2) \right. \nonumber \\
&-&2 \left.\dfrac{B \omega(I)}{\hbar} \beta_m \beta_{30} \tanh{\dfrac{B \omega(I)}{\hbar} \beta_m \beta_{30}}\right ).
\label{potential_2}
\end{eqnarray}
The potential (\ref{potential_2}) is used for the numerical diagonalization of $H_I$ and for the calculation of the parity splitting as a function of angular momentum.

Using the  dimensionless variable $x=\beta_{30}/\beta_m(I)$ and the parameter $\xi$ defined in Eq.~(\ref{xi}), one can rewrite  the Hamiltonian $H_I$ and the potential energy $V_I$ in a convenient form 
\begin{eqnarray}
H_I&=& \hbar \omega(I) h (\xi_I), \nonumber \\
h(\xi)&=& -\dfrac{1}{2\xi^2}\dfrac{d^2}{d x^2}+ v_\xi(x), \nonumber \\
v_\xi(x)&=&  \dfrac{1}{2}(\xi^2-1)+ \dfrac{1}{2}\xi^2 x^2 - \xi^2 x \tanh{(\xi^2 x)}.
\label{hamiltonian_1}
\end{eqnarray}
From Eq.~(\ref{hamiltonian_1}) it follows that the parity splitting can be parametrized as
\begin{eqnarray}
\Delta E(I)\equiv E_I^{(-)}-E_I^{(+)}=\hbar \omega(I) f \left ( \xi_I \right ),
\label{splitting}
\end{eqnarray}
where $ f \left ( \xi_I \right )$ is the energy of the first-excited state of the Hamiltonian $h(\xi)$ (the ground state energy of this Hamiltonian is zero). All  nucleus specific information  is contained in the dependence of $\xi_I$ on angular momentum and enters the Hamiltonian $h_\xi$ and the function $f(\xi_I)$ implicitly. Due to its universal character, it make sense to find an approximate analytical expression for $f(\xi_I)$. 

For small values of angular momentum ($\xi \ll 1$), the potential energy $v_\xi(x)$ reduces to that of an oscillator
\begin{eqnarray}
v_\xi(x)&=& \frac{1}{2}(\xi^2-1)+\frac{1}{2}(1-2 \xi^2)\xi^2 x^2, \\
&& \xi \ll 1.  \nonumber
\label{asym1}
\end{eqnarray}
The energy of the first excited state is then given by the frequency of an oscillator and we have
\begin{eqnarray}
f(\xi)=1- \xi^2.
\label{as1}
\end{eqnarray}

For large values of angular momentum ($\xi \gg 1$), $v_\xi(x)$ has a form of two oscillators separated by a large barrier
\begin{eqnarray}
\label{rasym2a}
v_\xi(x)&=& \frac{1}{2}(\xi^2-1)+\frac{1}{2}\xi^2 (|x|-1)^2, \\
&& \xi \gg 1.  \nonumber
\end{eqnarray}
The value of the energy interval between the two lowest levels of the double well potential for large barrier is given as \cite{Merzbacher}
\begin{eqnarray}
f(\xi)=\frac{2}{\sqrt{\pi}}\xi \exp{(-\xi^2)}.
\label{as2}
\end{eqnarray}

Both limits,  (\ref{as1}) and (\ref{as2}), are reproduced by one general expression
\begin{eqnarray}
f  (\xi )&=&\frac{\xi^2 e^{-\xi^2}}{2 \left [1+(1-e^{-\alpha \xi^2})\frac{\sqrt{\pi}}{4}\xi \right ]}  \coth\left(\frac  {\xi^2}{2} \right ), \nonumber \\
&&  \text{for} \hspace{10pt} \alpha=0.053, 
\label{approx_splitting}
\end{eqnarray}
where the value of the parameter $\alpha$ has been obtained by  fitting  the numerical results for the $f(\xi)$. The results obtained by numerical diagonalization of the Hamiltonian and those given by (\ref{approx_splitting}) are presented in Fig.~1.
In the limiting cases of very small and very large $\xi$ difference between the approximate and exact values of $f$ is negligible and vanishes asymptotically. The maximum deviation reaches about 2$\%$ at $\xi \approx 2$.

It is worth to note, that for the actual description of the experimental data, one can set $\alpha=0$, which yields simpler expression for the parity splitting as
\begin{eqnarray}
\Delta E(I)=\hbar \omega (I)\xi_I^2 e^{-\xi_I^2}  \coth\left(\xi_I^2/2 \right ).
\label{approx_splitting_1}
\end{eqnarray}
Since in the region of $\xi \gg 1$, the values of parity splitting are small and are influenced by many effects which are not included in the model (as for example, band crossing), we can neglect the deviations of (\ref{approx_splitting_1})  from (\ref{approx_splitting}) at large values of $\xi$ and use the expression (\ref{approx_splitting_1}).

\subsection{Dipole transitions}

In addition to the appearance of the low-lying negative parity states, a common property of nuclei exhibiting strong reflection-asymmetric correlations is the large values of the electric dipole transition probabilities \cite{Butler1996}.  While the absolute values of dipole moment for the transitions between negative- and positive-parity states
depend on nucleus,  its angular momentum dependence can be described, as it is shown below, by the universal function, similar to the situation with the parity splitting.   

\begin{center}
\includegraphics[width=7cm]{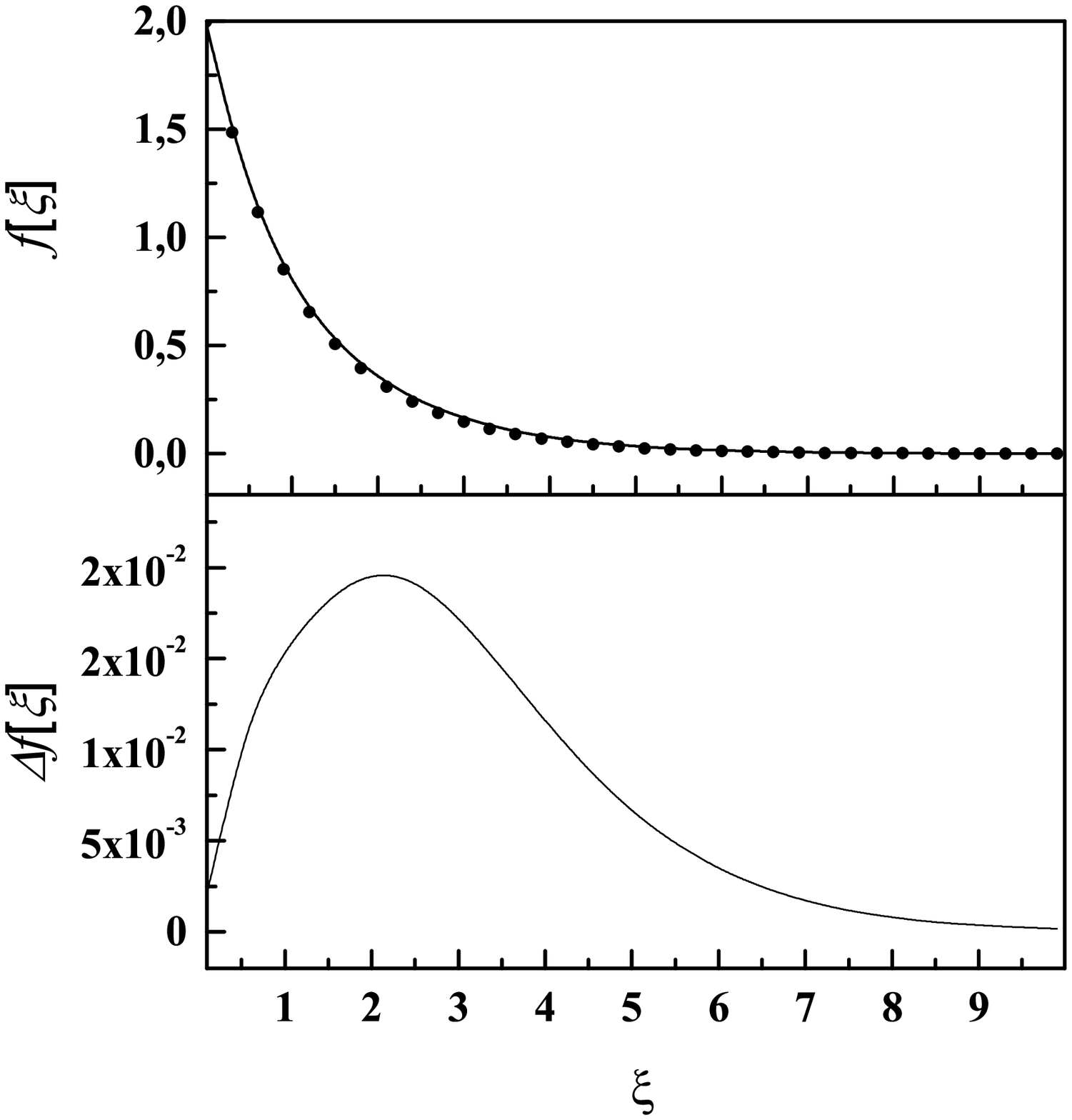}
\figcaption{\label{fig1}   Upper part: The function $f(\xi)$ obtained by the numerical diagonalization of the Hamiltonian $h_I$ (dots) and with an approximation of Eq. (\ref{approx_splitting}) (solid line). Lower part: difference between an  exact and approximated values of function $f(\xi)$. }
\end{center}

In the case of the well-deformed axially-symmetric nuclei, the operator of the  collective electric dipole moment can be written  in the intrinsic system as 
\begin{eqnarray}
D_0 \sim C \beta_{20} \beta_{30},
\label{dipole}
\end{eqnarray}
 where $C$ is the dipole polarizability  determined by the asymmetry between neutron and proton densities \cite{Butler1996}.   In the macroscopic liquid drop model, for example, we have 
\begin{eqnarray}
D_0 = C_{LD} A Z e \beta_{20} \beta_{30},
\label{dipole2}
\end{eqnarray}
where $ C_{LD}=0.0007$ fm \cite{Strutinsky1957}. 

Therefore, we see that the angular momentum dependence of the transitional dipole moment is determined by the matrix element $\langle i||\beta_{30}||f\rangle$,  where vectors $|i\rangle $ and $|f\rangle $ denote the initial and final states, respectively.  The lowest negative parity eigenfunction of the Hamiltonian (\ref{hamiltonian}) can be found numerically by the solution of the Schr\"odinger equation with potential (\ref{potential_2}). However, having the aim to get the result in the analytical form 
we supplemented the ansatz (\ref{ansatz}) for the positive parity ground state wave function by the expression  for the lowest negative parity wave function as
\end{multicols}
\begin{eqnarray}
\Psi^{(-)}_I(\beta_{30})&=&\left [ \dfrac{B \omega(I)}{\pi \hbar}\right ]^{1/4} \left ( 2 \left \{ 1-\exp{\left [-\dfrac{B \omega(I)}{\hbar}\beta_m^2 \right ]}\right \}\right )^{-1/2} \nonumber \\
&\times& \left ( \exp{\left [-\dfrac{B \omega(I)}{2\hbar}(\beta_{30}-\beta_m)^2 \right ]} - \exp{\left [-\dfrac{B \omega(I)}{2\hbar}(\beta_{30}+\beta_m)^2 \right ]}\right ).
\label{ansatz_1}
\end{eqnarray}
\begin{multicols}{2}
This form is confirmed by the numerical calculation. Note, that previously, in \cite{Jolos2011}, the methods based on supersymmetric quantum mechanics have been used.  Here, we use simpler way to reach an approximate expression for the parity splitting.
Using the expression for the parameter $\xi$ and a dimensionless variable $x$, the wave functions $\Psi^{(\pm)}(x)$ can be rewritten as
\end{multicols}
\begin{eqnarray}
\Psi^{(\pm)}(x)=\dfrac{\xi^{1/2}}{\beta_m^{1/2}\pi^{1/4}} \dfrac{1}{2\sqrt{1 \pm \exp{(-\xi^2)}}}
\times \left (\exp{\left [-\dfrac{1}{2}\xi^2 (x+1)^2\right ]}\pm\exp{\left [-\dfrac{1}{2}\xi^2 (x-1)^2\right ]} \right ).
\label{wf}
\end{eqnarray}
\begin{multicols}{2}

Using the ansatz (\ref{wf}) for the intrinsic wave functions of the positive and negative parity members of the alternating parity bands we can find an analytical expression for the angular momentum dependence of the matrix element of $\beta_{30}$, namely
\begin{eqnarray}
\langle i||\beta_{30}||f \rangle &=&\sqrt{\dfrac{\hbar}{B \omega}} \frac
{\sqrt{2}e^{\frac{(\xi_f^2-\xi_i^2)^2}{2 (\xi_f^2+\xi_i^2)}}(\xi_f \xi_i)^{1/2}
}
{\sqrt{(e^{\xi_f^2}-1  )  (e^{\xi_i^2}+1  )}} \nonumber \\
&\times& \dfrac{\left ( \xi_f^2-\xi_i^2+e^{\frac{2 \xi_f^2 \xi_i^2}{\xi_f^2+\xi_i^2}}(\xi_f^2+\xi_i^2) \right )}{(\xi_f^2+\xi_i^2)^{3/2}}.
\label{dipole_full}
\end{eqnarray}

The last expression can be simplified if we assumed that approximately $\xi_i \approx \xi_f=\xi$. Then
\begin{eqnarray}
\langle i||\beta_{30}||f\rangle=\sqrt{\frac{\hbar}{B \omega}}  \frac{\xi e^{\xi^2}}{\sqrt{e^{2\xi^2}-1}}.
\label{dipole_simple}
\end{eqnarray}

From  Eq.~(\ref{dipole_simple}) we see, that in the vibrational limit of the octupole motion ($\xi \ll 1$), we have 
\begin{eqnarray}
\langle i||\beta_{30}||f\rangle \approx \sqrt{\frac{\hbar}{B \omega}} .
\end{eqnarray}
For large values of $\xi$, the dipole moment is an increasing function of $\xi$. This increase is almost linear for $\xi>1$ 
\begin{eqnarray}
\langle i|\beta_{30}|f\rangle &\approx& \beta_m, \hspace{30pt} (\xi \gg 1).
\label{limits1}
\end{eqnarray} 
The angular momentum dependence of the dipole reduced transition probability  has the form
\begin{eqnarray}
B(E1,i \rightarrow f)= B(E1, 0^+ \rightarrow 1^-)  \frac{\xi^2 e^{2\xi^2}}{e^{2\xi^2}-1},
\label{dipole_angular}
\end{eqnarray}
where $\xi = \sqrt{\xi_i \xi_f}$.

\section{Results of calculations}

From Eq.~(\ref{splitting}) it follows, that the angular momentum dependence of the parity splitting is determined by the  function $f[\xi(I)]$. All information on the nucleus is contained in actual  dependence of $\xi$ on angular momentum, while the function $f[\xi]$ is universal. This function can be obtained numerically
   as the energy of the first excited state of the Hamiltonian $h_\xi$. Approximately, $f[\xi]$ is given by Eq.~(\ref{approx_splitting}). Therefore, in the following we use the function $f[\xi]$ given by Eq. (\ref{approx_splitting}) to describe the parity splitting of the nuclei.

Our calculations have shown that with sufficiently good accuracy the angular momentum dependence of $\hbar \omega_I$ and $\xi_I$ can be fitted as
\begin{eqnarray}
\hbar \omega_I &=& {\rm const}, \nonumber \\
\xi(I) &=& c I.
\label{parameters}
\end{eqnarray}
This parametrization contains very small number of parameters.
If we use this parametrization of the potential, we can see that the value of the frequency $\hbar \omega$ is immediately determined by the  value of the parity splitting at zero angular momentum, $\Delta E_{\textrm{exp}}(0)$. Indeed, if $I=0$ then   $\xi(0)=0$ and  $V_I(\beta_{30})$ reduces to the oscillator potential. Interval between the ground state and first excited state is then given by the frequency $\hbar \omega$.  Therefore, we obtain that $\hbar \omega =\Delta E_{\textrm{exp}}(0)$. The function $f \left [ \xi(I) \right ]$ is a universal function of $\xi$ and thus depends only on the parameter $c$ defined in Eq.~(\ref{parameters}). Moreover, if we use the results of \cite{Jolos2012}, we can connect the value of $c$ to the value of the critical angular momentum $I_{\textrm{crit}}$, at which  the phase transition from the octupole nondeformed to the octupole deformed shape takes place, namely
\begin{eqnarray}
 c I_{\textrm{crit}}=\dfrac{1}{\sqrt{2}}.
\label{crit}
\end{eqnarray} 

Finally, we obtain 
\begin{eqnarray}
\Delta E(I)=\Delta E_{\textrm{exp}}(0) f \left [ \dfrac{I}{\sqrt{2}I_{\textrm{crit}}}\right ].
\label{splitting1}
\end{eqnarray}

The choice of angular momentum dependence of $\xi$, given by (\ref{parameters}), can be qualitatively justified in the following way. At low angular momentum ($I < I_{\textrm{crit}}$), we can consider the alternating parity band as formed of two distinct bands consisting of even-parity and odd-parity states, respectively.  Defining the moments of inertia of positive- (negative-) parity bands as $\Im_{e}$($\Im_{o}$), the parity splitting can be obtained as
\begin{eqnarray}
\Delta E(I) &=& \Delta E(0)+\dfrac{\hbar^2 I(I+1)}{2 \Im_{o}(I)}-\dfrac{\hbar^2 I(I+1)}{2 \Im_{e}(I)} \nonumber \\
&=& \Delta E(0)-\dfrac{\hbar^2 I(I+1)}{2 \tilde \Im (I)}.
\label{momin1}
\end{eqnarray}
where
\begin{eqnarray}
\tilde\Im (I) = \dfrac{\Im_e(I) \Im_{o}(I)}{\Im_{o}(I)-\Im_e(I)}.
\end{eqnarray} 
At low $I$, we have for the moment of inertia of the positive parity states $\Im_{e}(I) \approx \Im (\beta_{30}=0)$.  The moment of inertia of the negative parity state $\Im_{o}(I)$ is a weakly depending  function of the angular momentum \cite{Shneidman2015}.
Comparing  (\ref{momin1}) with the approximated expression obtained using (\ref{as1}), we obtain
\begin{eqnarray}
\Delta E(0) = \hbar \omega, \nonumber \\
\hbar \omega \xi^2(I)=\dfrac{\hbar^2 I(I+1)}{2 \tilde \Im (I)}.
\label{11}
\end{eqnarray}
Since $\Im_{o}$ and, therefore, $\tilde \Im$ are weakly depending functions of $I$, the expression (\ref{11}) is in agreement with the approximation (\ref{parameters}).

At the limit of large angular momenta ($I  \gg I_{\textrm{crit}}$) the nucleus approaches the static octupole deformation and the  assumption of two separate rotational bands for positive- and negative-parity states is not valid anymore. In this limit nuclear  potential energy surface as a function of $\beta_{30}$ has  two pronounced minima separated by the barrier [see Eq.~(\ref{rasym2a})]. The parity splitting can then be determined as  \cite{Merzbacher}
\begin{eqnarray}
\Delta E(I)= \frac{2 \omega}{\sqrt{\pi}} \sqrt{\frac{2 V_0}{\hbar \omega }} \exp{\left ( -\frac{2 V_B}{\hbar \omega} \right  )},
\label{1}
\end{eqnarray}
where $V_B$ is the barrier between the right and left octupole minima. This barrier arises due to the fact that the nuclear moment of inertia increases with $\beta_{30}$. Since it is assumed that at $I=0$ the potential has a form of an oscillator [i.e. $V_B(I=0)=0$],  the barrier height can be determined as the 
difference between the rotational energies associated with change in the moment of inertia with $\beta_{30}$
\begin{eqnarray}
V_B(I) =\dfrac{\hbar^2 I(I+1)}{2 \Im(\beta_{30}=\beta_m)}-\dfrac{\hbar^2 I(I+1)}{2 \Im(\beta_{30}=0)} .
\end{eqnarray}

At large angular momentum the moment of inertia of negative and positive parity states are both close to the value at the minimum of the potential, i.e.  $ \Im(\beta_{30}=\beta_m) \approx \Im_o$. Therefore, we have
\begin{eqnarray}
V_B(I) =\dfrac{\hbar^2 I(I+1)}{2\tilde  \Im(I)} .
\label{barrier}
\end{eqnarray}
Comparing the expression (\ref{1}) with the barrier height in the form (\ref{barrier}) with the expression (\ref{as2}) we obtain
\begin{eqnarray}
\Delta E(0) = \hbar \omega, \nonumber \\
\hbar \omega \xi^2(I)=\dfrac{\hbar^2 I(I+1)}{\tilde \Im (I)},
\label{fff}
\end{eqnarray}
where the constancy of $\hbar \omega$ is again taken into account. 

Since $\tilde \Im$  is a weakly depending function of $I$,  the assumption (\ref{parameters}) is approximately valid in both limits of small and large angular momomenta. With the help of (\ref{parameters}),  (\ref{crit}) and (\ref{11}), the critical value of the angular momenta can be related to the change in the rotational energy caused by the mass asymmetric deformation dependence of the moment of inertia
\begin{eqnarray}
I_{\mathrm{crit}}= \gamma  \left ( \frac{\Delta E(0)}{2\hbar^2}  \dfrac{\Im_0 \Im_e}{\Im_0 - \Im_e} \right )^{1/2},
\label{critical}
\end{eqnarray}
where $\gamma$ is a constant close to unity. In this expression, the moment of inertia of the positive parity states should be calculated at small angular momenta, namely $\Im_e=\Im_e(I=2)$, while the moment of inertia of the negative parity states should be taken in the vicinity of the critical angular 
momentum $\Im_e=\Im_e(I=I_{\mathrm{crit}})$.

The calculation of the parity splitting is performed with use of the expression  (\ref{splitting1}) with the function $f$ in the form (\ref{approx_splitting}).  The experimental values of the parity splitting $\Delta E(I )$ are determined using the experimental energies $E_{\textrm{exp}}(I)$ of the lowest negative parity states and the positive parity states of the ground state band \cite{Data}.
 The quantity  $\Delta E(I )$ is determined as a difference between the energies of the negative- and the
positive-parity states with the same spin $I$. However, as it is described above, since at every value of $I$  there exist only one state with the
fixed parity $\pi=(-1)^I $,  the energy of the state of the opposite parity but with the same $I$ can be determined only by
interpolation using the energies of the states neighboring to $I$.  This interpolation should account for the angular momentum dependence of the excitation energy in the vicinity of $I$. Since in the model it is assumed that nuclei have  a stable
quadrupole deformation, the rotational law can be used, which leads to the following interpolation \cite{Jolos1995}
\begin{eqnarray}
E_{\mathrm{inter}}(I+1)=\frac{1}{2} [E_{\mathrm{exp}}(I+2)+E_{\mathrm{exp}}(I)] \nonumber \\
- \frac{1}{8} [E_{\mathrm{exp}}(I+4)-2  E_{\mathrm{exp}}(I+2)+E_{\mathrm{exp}}(I)]
\label{interpolation}
\end{eqnarray}
and the parity splitting is given by
\begin{eqnarray}
\Delta E(I)_{\mathrm{exp}} = (-1)^I (E_{\mathrm{inter}}(I)-E_{\mathrm{exp}}(I)).
\label{pspling}
\end{eqnarray}
Alternative expression for the parity splitting is given in \cite{Jolos2015}.  Both definitions produce almost identical numerical results for the parity splitting.

\end{multicols}
\begin{center}
\includegraphics[width=11.9cm]{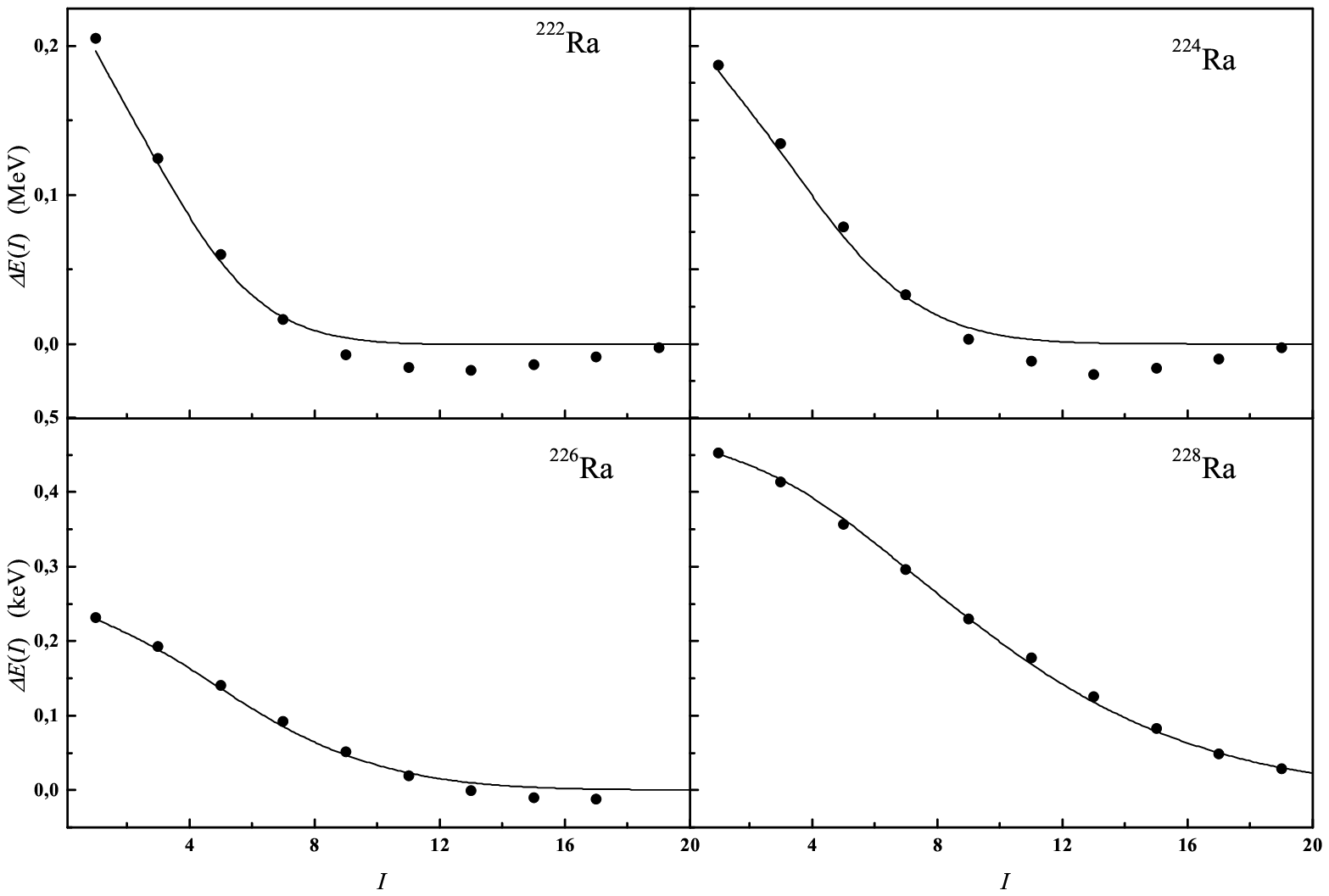}
\figcaption{\label{fig2}  Parity splitting as a function of angular momentum for various Ra isotopes.  Experimental data (circles) are taken from \cite{Data}. The calculated parity splittings (lines) are obtained as in Eq. (\ref{splitting1}) with use of the approximation (\ref{approx_splitting}). The values of the parameters $I_{\textrm{crit}}$ and $\Delta E(0)$ are given in Table~ \ref{tab:momcrit}.}
\end{center}

\begin{center}
\includegraphics[width=11.9cm]{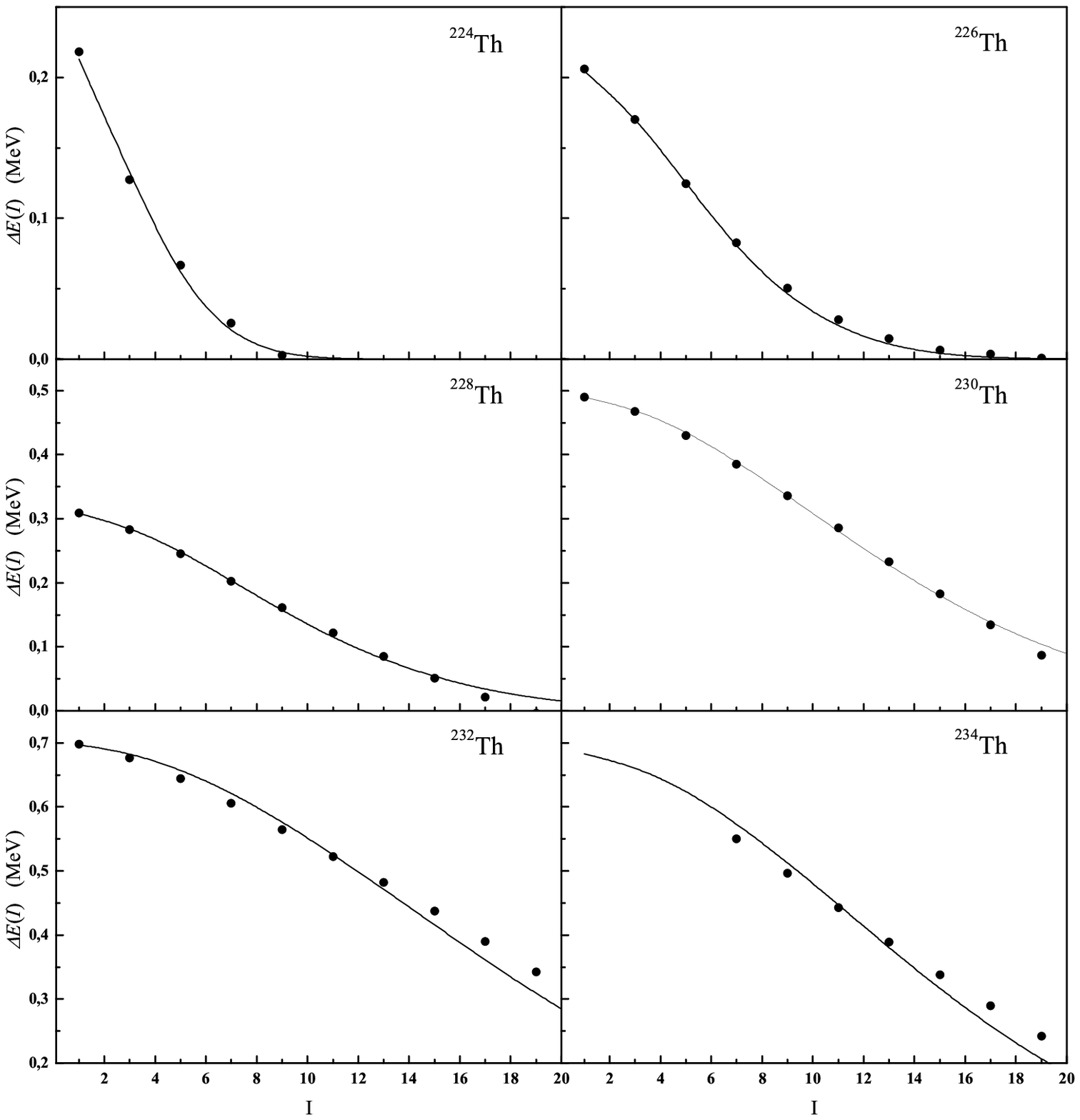}
\figcaption{\label{fig3}   Parity splitting as a function of angular momentum for various Th isotopes.  Experimental data (circles) are taken from \cite{Data}. The calculated parity splittings (lines) are obtained as in Eq. (\ref{splitting1}) with use of the approximation (\ref{approx_splitting}). The values of the parameters $I_{\textrm{crit}}$ and $\Delta E(0)$ are given in Table~ \ref{tab:momcrit}. }
\end{center}

\begin{center}
\includegraphics[width=11cm]{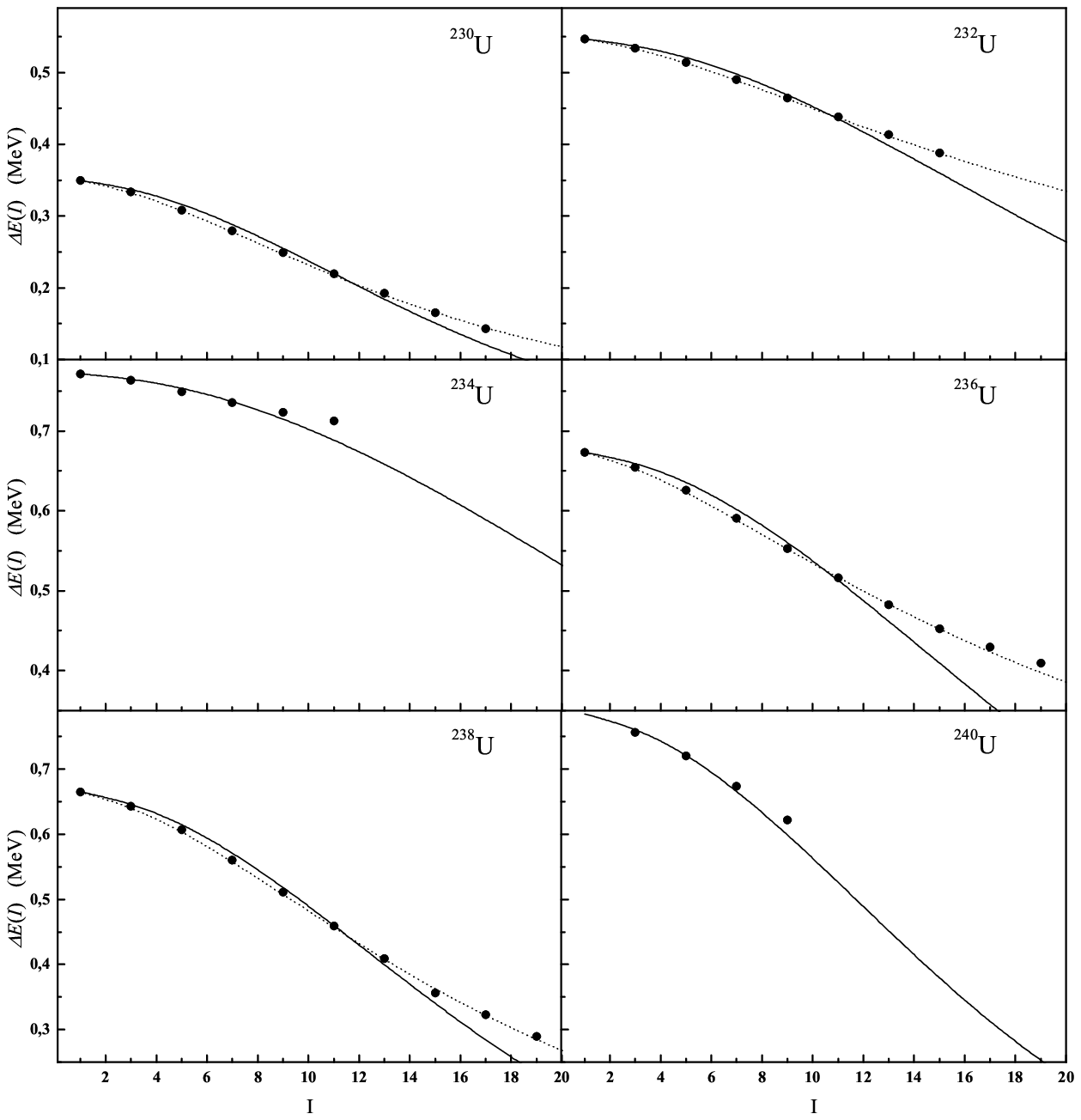}
\figcaption{\label{fig3}   Parity splitting as a function of angular momentum for various U isotopes.  Experimental data (circles) are taken from \cite{Data}. The calculated parity splittings (lines) are obtained as in Eq. (\ref{splitting1}) with use of the approximation (\ref{approx_splitting}). The values of the parameters  $I_{\textrm{crit}}$ and $\Delta E(0)$ are given in Table~ \ref{tab:momcrit}. The calculations performed with the dependence of $\xi$ on angular momentum taken in the form (\ref{newpar})  despayed  by the dashed line (see discussion in the text).}
\end{center}

\begin{center}
\includegraphics[width=11cm]{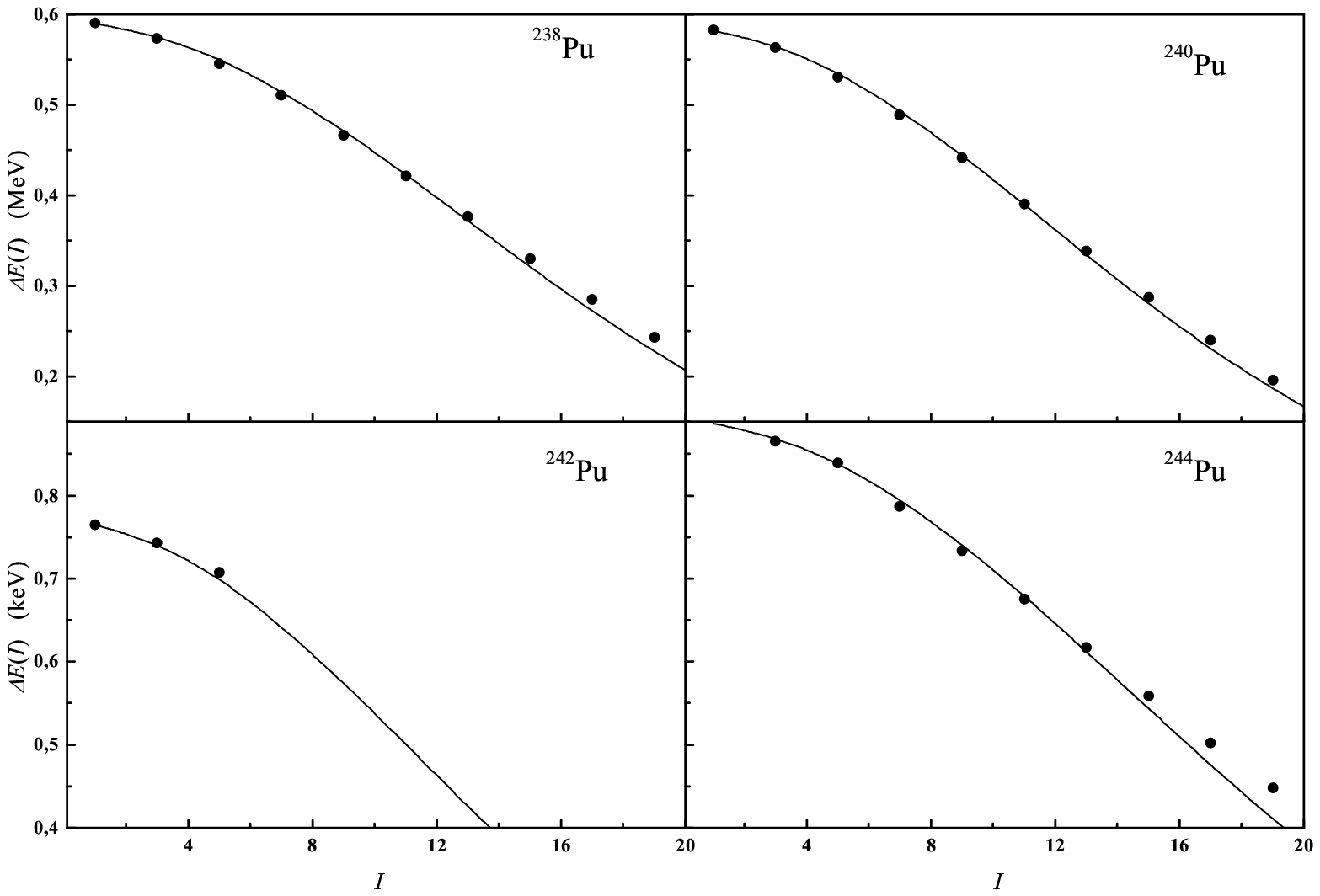}
\figcaption{\label{fig3}   Parity splitting as a function of angular momentum for various Pu isotopes.  Experimental data (circles) are taken from \cite{Data}. The calculated parity splittings (lines) are obtained as in Eq. (\ref{splitting1}) with use of the approximation (\ref{approx_splitting}). The values of the parameters $I_{\textrm{crit}}$ and $\Delta E(0)$ are given in Table~ \ref{tab:momcrit}.}
\end{center}

\begin{multicols}{2}

Since the experimental value of the parity splitting at $I=0$ is not available, the value $\Delta E(0)$ is fixed to reproduce experimental data for $\Delta E_{\mathrm{exp}}(1)$.
The critical angular momentum $I_{\mathrm{crit}}$ is fitted to give best overall description of the parity splitting in the range of angular momenta $0 \leq I \leq 20$. The reason why the larger values of the angular momenta is not considered is related to the possible appearance  of the band crossing at higher values of $I$. 
 The calculations performed for the deformed isotopes of Ra, Th, U, and Pu are presented on Figs.~{2-5} together with the experimental
data from \cite{Data}. The obtained values of the critical momenta $I_{\textrm{crit}}$ are presented in the Table~1. 

\end{multicols}

\begin{center}
\tabcaption{ \label{tab:momcrit}  The values of the parameters $\Delta E(0)$   (keV)  and $c$ used to describe  the parity splitting in the alternating parity bands of various actinide  are presented. Additionally, the last column contains the values of critical angular momenta $I_{\textrm{crit}}$ characterizing 
the phase  transition from octupole vibrations to the stable octupole deformation.}
\footnotesize
\begin{tabular*}{155mm}{@{\extracolsep{\fill}}|cccc||cccc|}
\hline 
 Nucleus &   $\Delta E(0)$   (keV)   &   $c$       & $I_{\mathrm{crit}}$ & Nucleus &  $\Delta E(0)$ (keV)   &   $c$  & $I_{\mathrm{crit}}$ \\
\hline
$^{222}$Ra &0.209  & 0.252     &2.81 &  $^{238}$Pu  &   0.584  & 0.053 & 13.32 \\
$^{224}$Ra &  0.192  & 0.210      & 3.37 & $^{240}$Pu  &  0.585  & 0.058 & 12.10 \\
$^{226}$Ra & 0.235 & 0.150      & 4.70 &  $^{242}$Pu  &  0.767  & 0.060   & 11.77 \\
$^{228}$Ra & 0.456  & 0.094     &7.53 & $^{244}$Pu  &  0.888  & 0.047 &   14.94 \\
\hline
$^{224}$Th & 0.226 & 0.247 & 2.86 &  $^{230}$U  &   0.351  & 0.063 &11.21     \\
$^{226}$Th &  0.209 & 0.149 & 4.88 & $^{232}$U  &   0.548  & 0.044 & 16.20\\
$^{228}$Th &  0.311  & 0.094 &7.54 &  $^{234}$U  &  0.772  & 0.031 &22.90 \\
$^{230}$Th & 0.492  & 0.069 &10.21&  $^{236}$U  &  0.674  & 0.046  &  15.39 \\
$^{232}$Th & 0.699  & 0.049 &14.50 &   $^{238}$U  &  0.669  & 0.056 &12.67\\
$^{234}$Th &  0.685  & 0.060  &11.77 & $^{240}$U  &  0.789  & 0.058 & 12.12\\
\hline

\end{tabular*}
\end{center}

\begin{multicols}{2}

We see a good overall agreement with experiment for all considered nuclei. There are discrepancies in behaviour of the calculated and experimental  dependencies of parity splitting which can be sorted 
in two ``groups''. The discrepancies of the first group are related to the fact that the experimental parity splitting can take negative values, while the calculated one approaches zero staying  positive. 
Among considered nuclei, this is the case for $^{222,224,226}$Ra and $^{224,226}$Th.  Such a behaviour of the parity splitting results from the coupling of the axially-symmetric octupole mode to the other modes which are not included into the model.
For example, all  nuclei in the considered mass region  have a negative parity band with $K=1$ \cite{Spieker2013}. This band can be interpreted as built on the excitation of non-axially symmetric octupole mode \cite{Shneidman2015}. The Coriolis coupling of this band
with the negative parity states of the alternating parity band shifts the latter down in  energy. Since there is no $\Delta K = 1$ partner band for the states of the positive parity, this perturbation  will decrease the parity splitting and, if unperturbed parity splitting is close to zero, will shift it  to the negative values.  

The effect of Coriolis coupling with the non-axially symmetric modes can only be seen in nuclei with not too large critical angular momenta. Indeed, the parity splitting adopt negative values only for $^{222,224,226}$Ra, whose critical angular momenta $I_{\textrm{crit}}$ are 
2.81, 3.37, and 4.70, respectively and for $^{224,226}$Th ($I_{\textrm{crit}}$=2.86, 4.88) .  
If critical angular momentum is large, this effect is hidden by the discrepancies of the second group. The second group combines the nuclei with large critical angular momenta, such as heavy Th isotopes and most of the considered U, and Pu isotopes (see Fig. 3-5). For these nuclei,
we see that the calculated parity splitting for $I>I_{\textrm{crit}}$ demostrates steeper incline than the experimental one. These discrepancies can be related to the centrifugal stretching.  Indeed, from Eq. (\ref{fff}) it follows that the linear dipendence of $\xi(I)$ given by (\ref{parameters}) can only 
be assumed  if the reduced moment of inertia does not depends on angular momentum.  This is, obviously, rather crude approximation for large values of $I$.  To improve the agreement with experiment at large values of $I$, we can assume
\begin{eqnarray}
\xi(I) = c I/(1+d I)
\label{newpar}
\end{eqnarray}
 instead of (\ref{parameters}). As an example, the effect of this additional term is demonstrated by the dashed line in Fig.~4 for $^{230,232,236,238}$U, whose alternating parity bands are long enough to account for the additional term in angular momentum dependence of $\xi$. As we see,  by adding the  parameter $d$,  agreement between  the calculated and expeirmental parity splittings at large angular momenta is improved. However, to keep the model simple, we avoid introduction of an additional parameter $d$.


Using the obtained values of the critical angular momenta $I_{\textrm{crit}}$, one can calculate the angular momentum dependence of the electric dipole transitional moment. In Fig.~6, the results for the $^{240}$Pu are presented. The calculated values are compared with the experimental data on dipole  moment   obtained in \cite{Wiedenhover1999}. In order to get the   values of  $D_0$ from the experimental data, we have assumed the stable quadrupole deformation and  an axial shape of considered nuclei.

\begin{center}
\includegraphics[width=7.0cm]{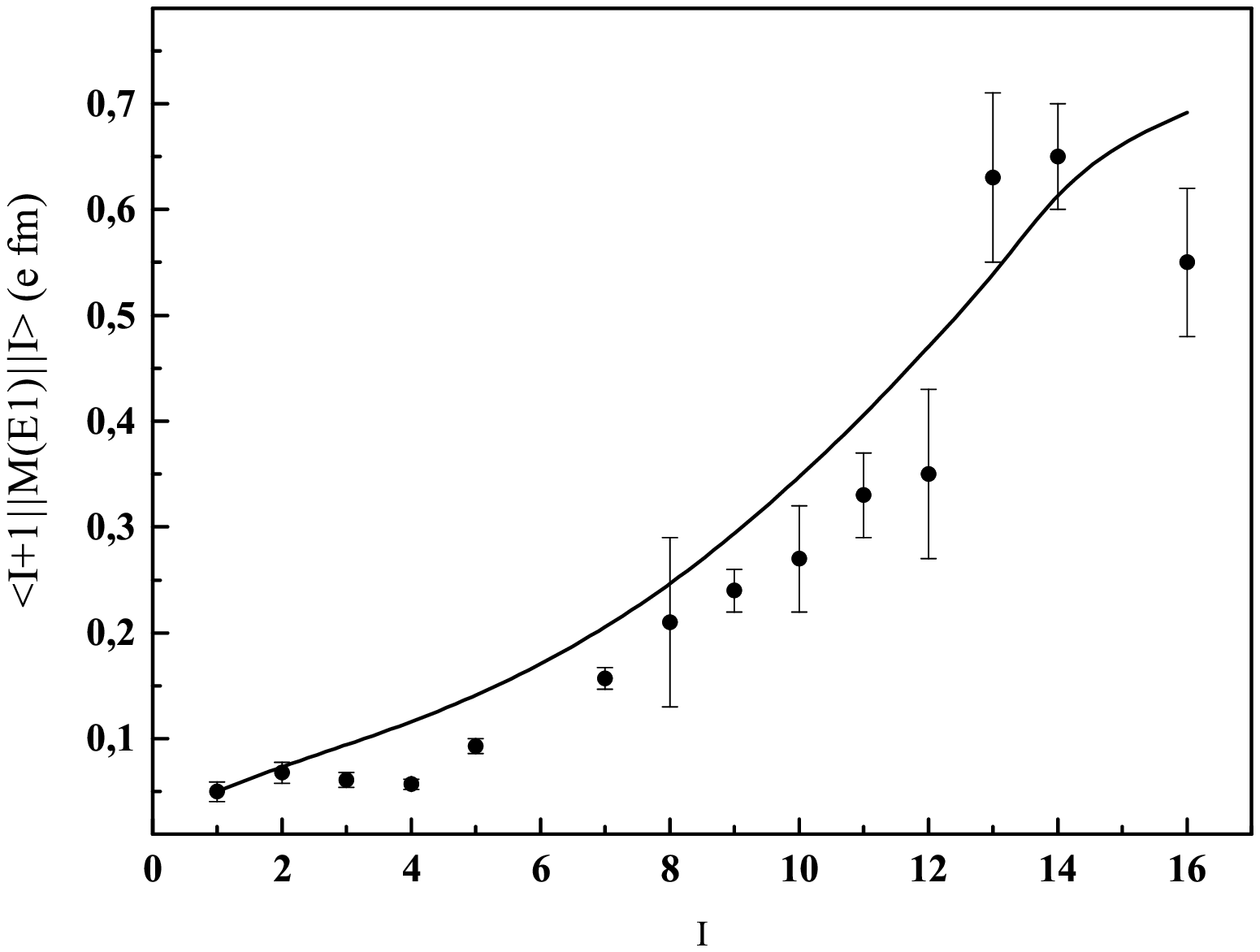}
\figcaption{\label{fig6}  Dependence of the calculated and experimental values of the transitional dipole moment on the angular momentum obtained for $^{240}$Pu. The calculations are performed with use of the expressions (\ref{dipole_angular}) and (\ref{trans}). The value of the critical angular momentum is given in Table~ \ref{tab:momcrit}.}
\end{center}

\begin{center}
\includegraphics[width=7.0cm]{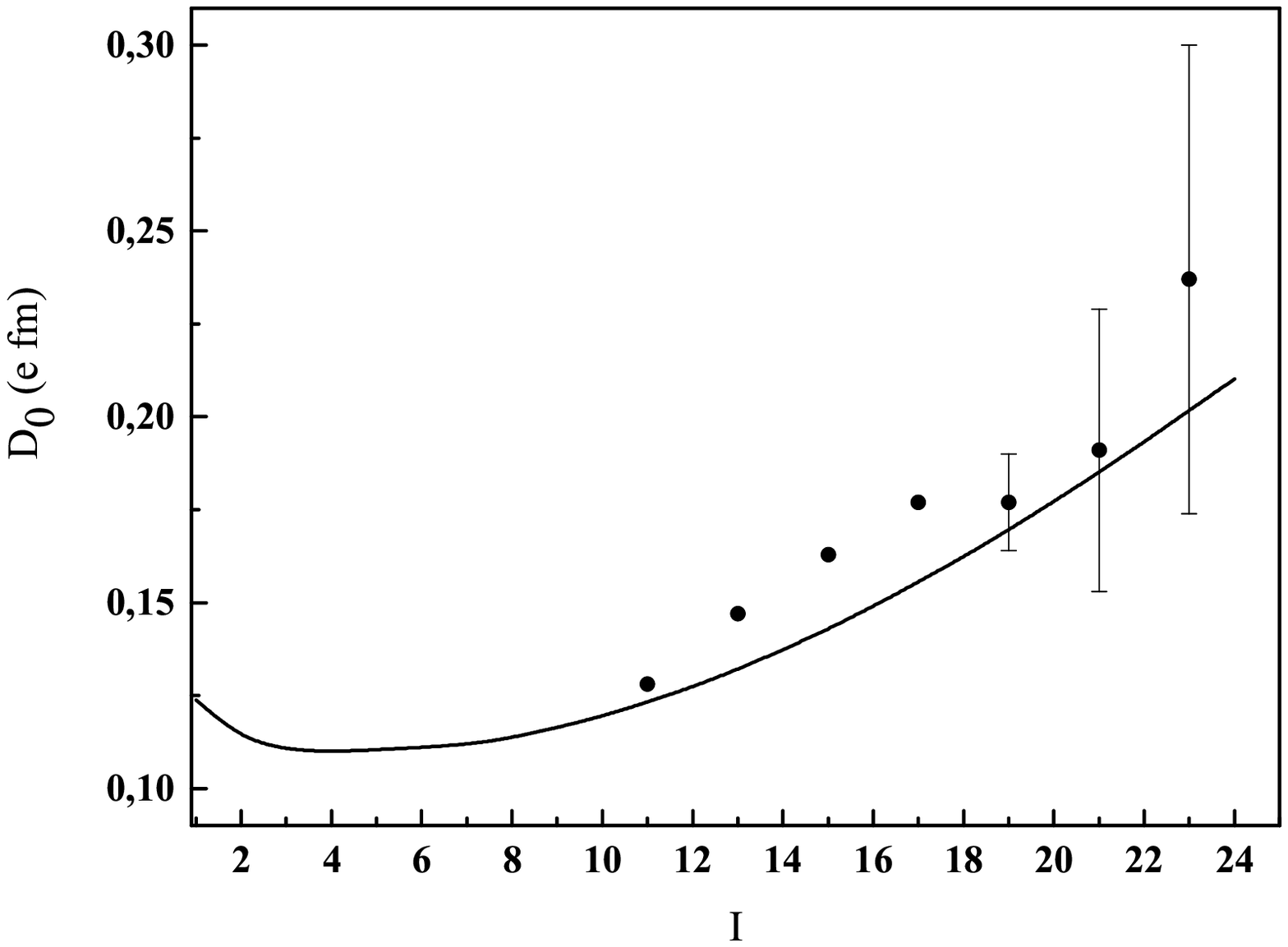}
\figcaption{\label{fig7}  Dependence of the calculated and experimental values of the $E1$ matrix element  on the angular momentum  for $^{226}$Ra. The calculations are performed with use of the expressions (\ref{dipole_angular}) and (\ref{trans}). The value of the critical angular momentum is given in Table~ \ref{tab:momcrit}.}
\end{center}

The dipole moment is obtained from the reduced transition probabilities using the expression \cite{Butler1996}
\begin{eqnarray}
B(E1,I \rightarrow I')&=&\frac{3}{4\pi}D_0^2 \left ( C_{I 0 \ 1 0}^{I' 0} \right )^2 \nonumber \\
&=&\frac{1}{2 I+1}\langle I'||M(E1)||I \rangle^2.
\label{trans}
\end{eqnarray}
As it  follows from  (\ref{dipole_angular}),  the absolute value of $D_0$ requires for its definition the experimental data on $B(E1,0^+\rightarrow 1^-)$. If these data are not available, the initial value of dipole-to-quadrupole moment is fitted to reproduce the lowest experimentally  available value of  $D_0/Q$. As it is seen from the results presented in Fig.~6, the calculation performed using Eq.~(\ref{dipole_angular}) and the critical angular momentum obtained from the fit of the parity splitting reproduces well the angular momentum dependence of the dipole transition probabilities along the alternating parity band. 

Another example is presented in Fig.~7 for the reduced matrix elements of the E1 transitions in $^{226}$Ra. The experimental data are taken from \cite{Wollersheim1993}. 
Unlike $^{240}$Pu which remains reflection symmetric until large angular momenta ($I_{\textrm{crit}}= 12.1$), the $^{226}$Ra has almost stable reflection-asymmetric deformation very close to the ground state ($I_{\textrm{crit}}= 4.7 $).  
As we see in both cases of small and large values of $I_{\textrm{crit}}$ the calculation with use of Eq. (\ref{dipole_angular}) agree well with the experimental data.

\section{Conclusion} 

A simple model proposed for the description of the spectroscopic information on the alternating parity bands in actinides. The model is based on the assumption that the yrast negative parity states of quadrupole-deformed nuclei are related to the excitation of an axially-symmetric octupole mode.
It is shown that the octupole deformation stabilizes with increase of  angular momentum, i.e., the phase transition occurs from octupole vibrations to the stable octupole deformation. This is caused by the dependence of the moment of inertia on the octupole deformation. As the moment of inertia increases with octupole deformation, its value is larger at the minimum of the potential than at the barrier height. As a result a depth of the deformation minimum increases with increase of angular momentum. 
For $^{222-228}$Ra,  $^{224-234}$Th, $^{230-240}$U, and $^{238-244}$Pu, the critical angular momenta $I_{\textrm{crit}}$ characterizing the phase transition to the reflection-asymmetric shape are calculated.  Relation of  $I_{\textrm{crit}}$ to the spectroscopic properties (such as the energies of the lowest
$I^\pi=1^-$ states and the moments of inertia of the positive parity and negative parity bands)  is obtained. 

Basing on this model the approximate analytical expressions for the angular momentum dependence of the parity splitting  (\ref{approx_splitting_1}) and the electric dipole transitional moment (\ref{dipole_angular}) are obtained.
These  analytical expressions contain a small number of the parameters with the clear physical meaning, namely, the frequency of the axially-symmetric octupole vibrations at zero angular momentum $\hbar \omega$  and the critical angular momentum  $I_{\textrm{crit}}$.
 These parameters can be fitted according to  the experimental data or calculated using a microscopical model.  We note that the same values of the parameters $\hbar \omega$ and $I_{\textrm{crit}}$ which are determined to get a good agreement with the
experimental data on parity splitting provide a good description of data on electric dipole transitional moment. The results obtained are illustrated by the calculations for different actinides  and compared  with the experimental data.

\end{multicols}

\clearpage
\end{CJK*}

\end{document}